\documentclass[prl,twocolumn,,amsmath,amssymb,floatfix]{revtex4}
\usepackage{graphicx}
\usepackage{bm}
\usepackage{amsmath,amssymb}

\def \be{\begin{equation}}
\def \ee{\end{equation}}
\def \ba{\begin{array}}
\def \ea{\end{array}}
\def \beq{\begin{eqnarray}}
\def \eeq{\end{eqnarray}}

\begin{document}
\title{Microscopic expression for the heat in the adiabatic basis.}

\author{Anatoli Polkovnikov$^1$}
\affiliation {$^1$Department of Physics, Boston University, Boston, MA 02215}

\begin{abstract}

We derive a microscopic expression for the instantaneous diagonal elements of the density matrix $\rho_{nn}(t)$ in the adiabatic basis for an arbitrary time dependent process in a closed Hamiltonian system. If the initial density matrix is stationary (diagonal) then this expression contains only squares of absolute values of matrix elements of the evolution operator, which can be interpreted as transition probabilities. We then derive the microscopic expression for the heat defined as the energy generated due to transitions between instantaneous energy levels. If the initial density matrix is passive (diagonal with $\rho_{nn}(0)$ monotonically decreasing with energy) then the heat is non-negative in agreement with basic expectations of thermodynamics. Our findings also can be used for systematic expansion of various observables around the adiabatic limit.

\end{abstract}
\maketitle

The standard definition of heat is the energy transferred from one body to another due to a temperature difference. However, from every day experience we know that we can generate a heat not only by transferring energy from say hot to cold object but by doing some mechanical work. For example, if we take a container with a gas isolated from the environment and then perform some cyclic process over this gas~\endnote{By a cyclic process we understand a process, where parameters of the Hamiltonian are allowed to change in time in an arbitrary way with the only restriction that at the end of the process the final Hamiltonian is equal to the initial one.}, e.g. compress and decompress it, then we will generate the heat unless we do it very slowly (adiabatically). In quantum mechanics for a cyclic process the heat will be equal to the energy increase caused by the transitions between different energy levels and thus it can be microscopically defined even for an isolated system. If two systems are brought to a contact for certain time and then again separated then heat can be defined in a similar fashion: energy change of a system due to transitions between different energy levels. On the other hand by work in thermodynamics one usually understands reversible part of the energy change, which is due to the change in the energy level structure but not due to the transitions. In a closed system then the total energy at each moment of time is
\be
\mathcal E(t)=\sum_{n} E_n(t)\rho_{nn}(t),
\ee
where $\rho_{nn}(t)$ are the diagonal elements of the density matrix in the instantaneous adiabatic basis. This expression can be rewritten as
\beq
&&\mathcal E(t)=\sum_n E_n(t)\rho_{nn}(0)+\sum_n E_n(t)(\rho_{nn}(t)-\rho_{nn}(0))\nonumber\\
&&~~\equiv \mathcal E_a(t)+Q(t) ,
\label{en}
\eeq
where $\rho_{nn}(0)$ are the diagonal matrix elements of the initial density matrix. The first term corresponds to the energy in the adiabatic process. Indeed as we know from quantum mechanics in the adiabatic limit there are no transitions between energy levels and thus diagonal elements of the density matrix do not change. The second term in Eq.~(\ref{en}) is related to the transitions between different levels and thus corresponds to the heat. Of course in large systems it is impossible to completely avoid transitions between energy levels~\cite{balian}. However, for slow processes the amount of states available for such transitions is small. Thus their contribution to the energy is negligible~\cite{np} and we recover the adiabatic limit. There is another reason why splitting of energy in Eq.~(\ref{en}) into the adiabatic part and the heat is consistent with thermodynamics. The first (adiabatic) term in this expression depends only on the state of the system, i.e. only on the initial state of the system and on the instantaneous parameters of the Hamiltonian. At the same time the heat is clearly a function of the process because it depends on the transitions between different energy levels during the evolution.

The heat, in turn, is intimately connected to the second law of thermodynamics (or simply the second law), which is the fundamental law of physics. In the most common formulation it states that the entropy of an isolated system can either increase or stay constant. However, there are other thermodynamically equivalent formulations of the second law~\cite{kardar}. In particular, the Kelvin's (Thompson's) formulation, used e.g. to prove impossibility of the {\em perpetuum mobile} of the second kind, states that it is impossible to create work by extracting heat from an isolated system. This statement implies that if we perform a cyclic process over some system then we can only increase its energy. Thus the heat in Eq.~(\ref{en}) should be nonnegative if we start from thermal equilibrium.   And indeed this statement was microscopically proven in Refs.~\cite{armen_kelv, chris}. In Ref.~\cite{armen_mwp} it was further shown that $Q(t)\geq 0$ for non-cyclic processes (and still initial thermal equilibrium state) as long as there are no unavoided level crossings between different energy levels. This statement is called the minimum work principle. In Ref.~\cite{thirring} a more general statement was formulated that the work on the system or equivalently heat is nonnegative for any cyclic process if the initial density matrix is passive, i.e. i) it is diagonal and ii) microscopic occupancies of different energy levels are nonincreasing functions of energy, i.e. $(\rho_n-\rho_m)(E_m-E_n)\geq 0$.

The main purpose of this paper is to find a microscopic expression for the heat in the system under the assumptions and relate it to many-body transition probabilities. The other purpose of this work is to define the framework for the perturbation theory around the adiabatic limit. For this purpose it is natural to work in the adiabatic or instantaneous basis. This basis diaogonalizes the Hamiltonian at each moment of time and thus if the Hamiltonian changes in time so does the basis. In Ref.~\cite{ap_adiabatic} it was already demonstrated how one can use this basis to calculate scaling properties of the density of excitations near critical point doing the first order perturbation theory in the rate of change of external parameters. In Ref.~\cite{claudia_sg} it was shown how using this perturbation theory in the adiabatic basis for a particular sine-Gordon model one can find the number of excitations and the heating using form-factors. Basically finding dynamical observables was reduced in that work to taking integrals over known static matrix elements.

Here we find a microscopic expression for the heat and more generally for diagonal matrix elements of the density matrix under the conditions that the initial density matrix is stationary (diagonal):
\be
\rho_{nn}(t)=\rho_n^0+\sum_m (\rho_m^0-\rho_n^0)|I_{nm}(t)|^2,
\label{dens}
\ee
where $\rho_n^0$ is the initial value of the occupancy of the $n$-th energy level, $I_{nm}(t)\equiv I_{nm}$ are the microscopically defined transition amplitudes, which will be explicitly defined below. Remarkably the sum in Eq.~(\ref{dens}) contains only squares of the absolute values of $I_{nm}$ suggesting that $|I_{nm}|^2$ can be interpreted as a transition probability from $m$-th to $n$-th state. Note that since $|I_{nn}|^2$ in Eq.~(\ref{dens}) is arbitrary one can always choose it such that $\sum_m |I_{nm}|^2=1$. As we point below the unitarity of the evolution requires that transition probabilities satisfy the sum rule:
\be
\sum_m |I_{mn}|^2=\sum_m |I_{nm}|^2.
\ee
In general $|I_{mn}|\neq |I_{nm}|$ though the transition probabilities are equal to each other within (i) the Fermi-Golden rule, which applies, for example, for infinitesimally short transition times, or (ii) for cyclic processes where the external parameter symmetrically changes in time. Using Eq.~(\ref{dens}) to find the total energy of the system and subtracting the adiabatic part we find
\be
Q=\sum_{m,n} E_n (\rho_m^0-\rho_n^0)|I_{nm}|^2=\sum_{m,n} \rho_m^0 (E_n-E_m)|I_{nm}|^2.
\label{heat}
\ee
For a cyclic process where final and initial energies are identical: $E_n=E_n^0$ one can show $Q(t)\geq 0$ if the initial state is passive~\cite{thirring}. If the transition probabilities satisfy detailed balance $|I_{mn}|^2=|I_{nm}|^2$ then the proof trivially follows from the symmetrization of the sum with respect to the indices $m$ and $n$. In general the proof is more subtle but the statement is still correct~\cite{armen_kelv, thirring}. From the explicit expressions below we find that in the adiabatic limit all $I_{nm}\to 0$ as long as $E_n\neq E_m$ and we recover $Q\to 0$. For non-cyclic process the heat is still positive under the same assumptions as long as there are no unavoided level crossings~\cite{armen_mwp}.

Let us now make a few remarks on Eqs.~(\ref{dens}) and (\ref{heat}):

\vspace{0.2cm}
\begin{itemize}

\item In derivation of Eq.~(\ref{dens}) it is important that the initial density matrix is diagonal only at the initial time, the following evolution can be arbitrary and the density matrix at any later time $t$ is not necessarily diagonal.

\item The Gibbs distribution $\rho_n\propto \exp(-\beta E_n)$ is obviously a passive state resulting in non-negative heat. However, the same statement equally applies to initial generalized Gibbs distribution functions suggested for integrable systems~\cite{olshanii_gen_gibbs}.

\item The corollary of the Kelvin's statement is also true: if one starts from an overheated configuration where higher energy states have higher occupancy then the heat after a cyclic will be non positive, i.e. it will not be possible to do a positive work on the overheated system.

\item The state where all energy levels are equally populated (and the initial density matrix is stationary) is neutral in a sense that its diagonal matrix elements can not be changed in any process. Correspondingly one can not add or extract the energy from this state.

\end{itemize}

In the remainder of the paper we will give the details of the derivation of Eq.~(\ref{dens}). First we will show how the derivation works for a two-level system and then we will generalize the result.

{\em Two-level system.} Suppose we are dealing with a Hamiltonian two level system. We will work with the Schr\'odinger equation for the wave function describing the system and then take the average over the initial density matrix. Alternatively one can directly work with the Liouville equation for the density matrix. In the adiabatic basis (co-moving with the Hamiltonian) the Schr\"dinger equation for the wave function takes the form~\cite{ap_adiabatic}:

\be
{d a_2\over dt}=-R_{2,1}(t) a_1(t),\quad {d a_1\over dt}=-R_{1,2}(t) a_2(t)\label{schr2},
\ee
where $a_1(t)$ and $a_2(t)$ are the coefficients of the expansion of the wave function in the adiabatic basis describing our two level system $\psi(t)=a_1(t)|1\rangle+ a_2(t) |2\rangle$. The states $|1\rangle$ and $|2\rangle$ are the eigenstates of the instantaneous Hamiltonian: $\mathcal H(t)|n\rangle =E_n(t) |n\rangle$, $n=1,2$. Finally $R_{2,1}(t)=-R^\star_{1,2}(t)$ is given by
\be
R_{2,1}(t)=\langle 2|\partial_t|1\rangle\exp\left(i\int_0^t (E_2(t')-E_1(t'))dt'\right).
\ee
For simplicity we omitted the Berry phase term here. The latter can be always reabsorbed into definition of $R_{2,1}$ and thus it is not going to affect our conclusions. We work in units where $\hbar=1$. The condition $R_{2,1}=-R_{1,2}^\star$ ensures the unitarity of the evolution of the wave function. Eqs.~(\ref{schr2}) can be rewritten in the integral form:
\beq
a_2(t)=a_2(0)-\int_0^t R_{2,1}(t') a_1(t') dt'\label{int1}\\
a_1(t)=a_1(0)+\int_0^t R_{2,1}^\star(t') a_1(t') dt'\label{int2}
\eeq
We are interested in the diagonal elements of the density matrix at time $t$ (by analogy one can also-derive off-diagonal matrix elements). For example
we have $\rho_{22}(t) =\overline{|a_2(t)|^2},$
where the overline implies averaging over the initial density matrix. A similar expression can be found for $\rho_{11}(t)$. Using Eqs.~(\ref{int1}) and (\ref{int2}) we find
\beq
&&\rho_{22}(t)=\rho_2^0+\int\limits_0^t\int\limits_0^t dt' dt'' R_{2,1}(t') R_{2,1}^\star(t'') \overline{a_1(t') a_1^\star(t'')}\nonumber\\
&&-\int\limits_0^t dt'\overline{a_2^\star(0) R_{2,1}(t') a_1(t')+ a_2(0)R_{2,1}^\star(t')a_1^\star(t')}.\phantom{X}
\label{ef0}
\eeq

Let us first prove our statement assuming that we are close to the adiabatic limit and the system is only slightly excited~\endnote{This assumption, however, does not imply that the Hamiltonian does not change significantly during the evolution and that the linear response approximation is valid.}.
Using the fact that $\overline{ a_2^\star(0) a_1(0)}=0$ we find
\beq
&&\rho_{22}(t)\approx \rho_2^0-\rho_2^0
\left[G_{2,1}\circ G_{2,1}^\star+G_{2,1}^\star\circ G_{2,1}\right]\nonumber\\
&&~+\rho_1^0 |G_{2,1}|^2=\rho_2^0+(\rho_1^0-\rho_2^0)|G_{2,1}|^2.
\label{ef1}
\eeq
Here $G_{n,m}$ and their convolutions are defined as follows:
\beq
&&G_{2,1}\equiv G_{2,1}(t)=\int_0^t R_{2,1}(t') dt',\nonumber\\
&&G_{2,1}\circ G_{2,1}^\star=\int_0^t dt' R_{2,1}(t')\int_0^{t'}  dt'' R_{2,1}(t'').
\eeq
Note that these convolutions can be also expressed through time-ordered products. To get from the first to the second line in Eq.~(\ref{ef1}) we used an obvious identity $G_{2,1}\circ G_{2,1}^\star+G_{2,1}^\star\circ G_{2,1}=|G_{2,1}|^2$. In the adiabatic limit (assuming that the system is not degenerate) we have $G_{2,1}=0$ and hence we have $\rho_{22}(t)=\rho_2^0$. In general $G_{2,1}\neq 0$ if the transitions between the two levels are not forbidden and as a result $\rho_{22}(t)> \rho_2^0$ as long as $\rho_1^0>\rho_2^0$, which is the condition of passivity for a two-level system.

Beyond the perturbation theory we can solve Eqs.~(\ref{int1}) and (\ref{int2}) by iterations and substitute the solution into Eq.~(\ref{ef0}). Since we give a general derivation for a multi-level system below we will only state the final result:
\beq
&&\rho_{22}(t)=\rho_2^0+(\rho_1^0-\rho_2^0)|I_{2,1}|^2,\\
\label{main2}
&&I_{2,1}={G_{2,1}\over 1+G_{2,1}^\star\circ G_{2,1}}.
\label{I21}
\eeq
The quantity $I_{2,1}$ is understood in terms of the Taylor expansion:
$I_{2,1}=G_{2,1}-G_{2,1}\circ G_{2,1}^\star\circ G_{2,1}+\dots$ Alternatively $I_{2,1}$ can be represented as a time ordered exponent (see Eq.~(\ref{I_roma}) below.

{\em Multi level generalization.} We can repeat now a similar derivation for a general system containing  multiple energy levels. We again will work in the adiabatic basis. Then instead of Eq.~(\ref{int1}) we find
\be
a_n(t)=a_n(0)-\sum_m' \int_0^t R_{n,m}(t') a_m(t') dt',
\label{int3}
\ee
where the prime over the summation index implies that the term with $m=n$ is excluded. For simplicity we again assume that there is no Berry phase appearing in the problem, the latter can be always reabsorbed into the definition of $R_{n,m}$. Using Eq.~(\ref{int3}) one immediately finds the generalization of Eq.~(\ref{ef0}):
\begin{widetext}
\be
\rho_{nn}(t)=\rho_n^0+\sum_{m,p} \int\limits_0^t\int\limits_0^t dt' dt'' R_{n,m}(t') R_{n,p}^\star(t'') \overline{a_m(t') a_p^\star(t'')}- \sum_{m} \int\limits_0^t dt'\,\overline{a_n^\star(0) R_{n,m}(t') a_m(t')+ a_n(0)R_{n,m}^\star(t')a_m^\star(t')}.
\label{efn}
\ee
\end{widetext}
Let us again assume that we are close to the adiabatic limit and do the perturbative expansion in $G_{n,m}=\int_0^t R_{n,m}(t') dt'$. In the leading order we find
\be
\rho_{nn}(t) =\rho_n^0+\sum_{m} (\rho_m^0-\rho_n^0) |G_{m,n}|^2.
\ee
In the adiabatic limit $G_{n,m}\to 0$ for all $n\neq m$ as long as $E_n\neq E_m$ and we recover $\rho_{nn}(t)=\rho_n^0$. If the process is not adiabatic and the transitions  between different levels are not forbidden then diagonal matrix elements of the density matrix change in time. The equation above can be generalized beyond the first order of the perturbation expansion. We can formally solve Eq.~(\ref{int3}):
\be
a_n(t)=a_n(0)-\sum_m I_{n,m} a_m(0),
\ee
where
\be
I_{n,m}=\left(G\over 1+G\right)_{n,m}.
\label{I_nm}
\ee
Note that formally $I_{nm}$ can be expressed through a time ordered exponent~\cite{roma}:
\be
{\bf I}=-T_\tau \exp\left[-\int_0^t d\tau {\bf R}(\tau)\right]+{\bf 1},
\label{I_roma}
\ee
where ${\bf 1}$ is the unity operator. Next we substitute this expression to Eq.~(\ref{efn}) to get:
\be
\rho_{nn} (t)=\rho_n^0+\sum_{m} \rho_m^0\left|I_{nm}\right|^2-\rho_n^0 \left(I_{nn}+I^\star_{nn}\right).
\label{ef3}
\ee

Now instead of checking the desired properties of the Green's functions leading to Eq.~(\ref{dens}) we will use the following trick. Consider the normalization of the wave function $\sum_n |a_n(t)|^2=1$. This implies
\be
1=1+\sum_{n,m} \rho_n^0\left| I_{mn}\right|^2-\sum_n \rho_n^0 \left(I_{nn}+I^\star_{nn}\right).
\label{trick}
\ee
 Clearly the normalization should be conserved for any choice of $\rho_n^0$, therefore
\be
I_{nn}+I^\star_{nn}=\sum_{m} \left| I_{mn}\right|^2=\sum_{m} \left| I_{nm}\right|^2.
\label{green}
\ee
The identity above actually follows from the unitarity of the evolution and it is equivalent to the optical theorem in scattering theory~\cite{LL3}. Note that in the sum above we are allowed to interchange indices $m$ and $n$ because
$(I^\dagger I)_{nn}=(II^\dagger)_{nn}$. This identity trivially follows from the fact that $I$ is the difference between unity and unitary operators (see Eq.~(\ref{I_roma})). Substituting Eq.~(\ref{green}) into Eq.~(\ref{ef3}) we recover Eq.~(\ref{dens}).

The result~(\ref{dens}) can be used for finding expectation values of other observables commuting with the Hamiltonian or of the time averaged values of arbitrary observables following a quench. If $\Omega$ is such an observable then
\be
\langle \overline{\Omega} \rangle = \sum_n \Omega_{nn}\rho_n^0+\sum_{m,n} \Omega_{nn} (\rho_m^0-\rho_n^0) |I_{nm}(t)|^2,
\label{a_f}
\ee
where overline implies the time averaging, which is necessary if $\Omega$ does not commute with the Hamiltonian. As long as the matrix elements of $\Omega$ satisfy the conditions of passivity: for $\rho_n\geq \rho_m$ we have $\Omega_{nn}\geq \Omega_{mm}$ ($\Omega_{nn}\leq \Omega_{mm})$ the the expectation value of this operator can either decrease (increase) in an arbitrary cyclic process. In particular, one can choose $\mathcal A$ to be equal to any power of the Hamiltonian an make similar statements about the moments of the Hamiltonian. We also note that Eq.~(\ref{a_f}) gives expectation values of steady state values of arbitrary stationary observables not commuting with the Hamiltonian. This is true because if the Hamiltonian does not change in time then all off-diagonal elements of the density matrix oscillate in time and thus average to zero once the system reaches the steady state.

The expansion (\ref{I_nm}) together with Eq.~(\ref{dens}) can be used to develop perturbation theory in the quench rate $\delta$ (or the rate of change of external parameters) near the adiabatic limit. In Refs.~\cite{ap_adiabatic, np, claudia_sg} such expansion in the leading order was used to analyze slow dynamics for particular situations. It was shown that this perturbation theory can be applied when the conventional perturbation theory in interaction strength diverges, for example, in low dimensional systems particularly near phase transitions. We also emphasize that Eq.~(\ref{dens}) was obtained under the assumption of the initially diagonal density matrix. If this is not true then the expression for $\rho_{nn}(t)$ will acquire additional terms linear in $\rho_{n,m}(0)$ for $n\neq m$. Those terms usually strongly fluctuate in the phase space in complex systems. Then these additional terms will average to zero and thus will not affect the Eq.~(\ref{dens}).

In conclusion we derived a microscopic expression for the diagonal elements of the density matrix~(\ref{dens}) in the instantaneous adiabatic basis provided that initial density matrix is stationary (diagonal).  This expression can be used to find explicit formulas for the heat, which is consistent with the Kelvin's formulation of the second law of thermodynamics. In the adiabatic basis all transition probabilities are expressed as squares of the matrix elements of the many-body evolution operator. The time only enters as an external parameter so there is no qualitative difference between time dependent and time-independent problems (if we are interested only in diagonal matrix elements of the density matrix). The Eq.~(\ref{dens}) can be used to construct an adiabatic perturbation theory, to perform Monte-Carlo simulations for time-dependent problems, and to derive general connections between various stationary observables in the out of equilibrium systems.

\acknowledgments The author is  grateful to H.~Spohn for pointing that in general $|I_{mn}\neq |I_{nm}|$ and correcting related mistakes in the earlier version of this manuscript. The author would like to acknowledge A.~Allahverdyan, R.~Barankov, A.~Castro Neto, S.~Girvin, V. Gurarie, D.~Huse, C.~ Jarzynski, Y.~Kafri, and W.~Zwerger for useful discussions. This work was supported by AFOSR YIP.

\end{document}